\documentclass[12pt]{article}
\usepackage{latexsym,amsfonts,pictex}

\newcommand{\text}[1]{\mbox{\rm #1}}
\newenvironment{romanlist}{\begin{list}{(\theenumiii)}
{\usecounter{enumiii}}}{\end{list}}
\newcommand{\D}{\mathcal{D}}
\newcommand{\E}{\mathcal{E}}
\newcommand{\F}{\mathcal{F}}
\newcommand{\G}{\mathcal{G}}
\newcommand{\M}{\mathcal{M}}
\newcommand{\Q}{\mathcal{Q}}
\newcommand{\Y}{\mathcal{Y}}
\newcommand{\x}{\mathbf{x}}

\newcommand{\be}{\mathbf{e}}
\newcommand{\bn}{\mathbf{n}}
\newcommand{\bs}{\mathbf{s}}
\newcommand{\bC}{\mathbf{C}}
\newcommand{\bK}{\mathbf{K}}
\newcommand{\bU}{\mathbf{U}}
\newcommand{\sgn}[1]{\mathrm{sgn~}#1}
\newcommand{\dom}[1]{\mathrm{dom~}#1}
\newcommand{\tr}[1]{\mathrm{tr~}#1}

\begin{document}
\title{The monodromy matrix method of solving an exterior boundary value
problem for a given stationary axisymmetric perfect fluid solution}
\author{I.\ Hauser\thanks{Home Address: 4500 19th Street, \#342, 
Boulder, CO 80304} and F.\ J.\ Ernst\thanks{Email: gravity@slic.com} \\
FJE Enterprises\thanks{Homepage URL: http://pages.slic.com/gravity},
511 CR 59, Potsdam, NY 13676}
\date{September 14, 2000}
\maketitle
\begin{abstract}
A procedure is described for matching a given stationary axisymmetric
perfect fluid solution to a not necessarily asymptotically flat vacuum 
exterior.  Using data on the zero pressure surface, the procedure yields 
the Ernst potential of the matching vacuum metric on the symmetry axis.  
From this the full metric can be constructed using a variety of well 
established procedures.
\end{abstract}


\section{Introduction} 
Because of the tremendous strides that were taken during the 1970's and 
1980's in coping with the mathematical problems presented by stationary 
axisymmetric vacuum (SAV) and electrovac fields, the attention of many 
workers in this field has shifted in recent years to those even more 
challenging problems that are associated with stationary axisymmetric 
perfect fluid (SAPF) spacetimes.  Except in the case of zero pressure
fluids (dust) there exists no general method for solving the Einstein
field equations within the fluid itself.  Neither has any general method 
for joining a given SAPF solution to a vacuum SAV exterior at a zero 
pressure surface been described.  It is the latter problem to which we 
shall propose a solution in this paper.  

From the outset we stress that we are {\em not\/} insisting that the SAV
metric be asymptotically flat.  Ours is a method with which to effect a
matching of a {\em given\/} SAPF to a SAV exterior, and not a method for 
solving a {\em global\/} problem.  In the latter case, the asymptotic 
flatness and the singularity free nature of the solution are built in 
ab initio, and the SAPF metric as well as the SAV metric are determined 
a posteriori.  The rotating dust disk solution of Neugebauer and 
Meinel\cite{dustdisk} remains the only known global solution of the 
stationary axisymmetric field equations corresponding to a physically 
interesting rotating source, the analog of the classical zero pressure 
MacLaurin disk. 

Restricting their attention to dust disks, Neugebauer and Meinel employed 
a method involving two integral equations, the ``small'' and ``big'' 
integral equations.  The former integral equation was derived from the 
Neugebauer linear system for the Ernst equation under the assumption that 
on the disk $0 \le \rho \le \rho_{0}$ the Ernst potential $\E(0,\rho)$ is 
real and independent of $\rho$, and under the assumption that there exists 
a singularity free asymptotically flat matching SAV metric.  The solution 
of this first integral equation yielded the axis values $\E(z,0)$ for $z>0$ 
of the Ernst potential.  The second integral equation was derived from a 
Riemann-Hilbert problem, and was used in order to obtain the Ernst potential 
$\E(z,\rho)$ of the vacuum metric that matched the axis data derived in the 
first step.  Finally, the surface density of the dust disk was determined
by computing the jump in one of the metric fields as the disk is crossed.

In the present paper, and with an eye toward providing a procedure that can
be employed routinely by those who discover new perfect fluid solutions, we 
develop the basic machinery that will be required to treat the ``joining
problem'' for a broad class of spinning fluid spheroids that will be defined
in Sec.\ \ref{mat}C.  Our approach, which is a variation 
of the ``monodromy transform'' approach of Alekseev\cite{monodromy}, employs 
matching data, essentially certain metric components and their differentials, 
on the zero pressure surface to compute a $2 \times 2$ matrix function 
$\Pi(\tau)$ of a complex spectral parameter $\tau$, the so-called 
``monodromy matrix.''  

The axis values of the Ernst potential of the exterior SAV solution can be 
determined immediately from $\Pi(\tau)$, which means that any of the standard
methods of generating a SAV solution from its axis data can be employed. 
Thus, for example, the SAV solution can be constructed by solving a 
homogeneous Hilbert problem (HHP) that was developed by the 
authors\cite{he1,he2,he3} to effect Kinnersley-Chitre\cite{KC} 
transformations. 

Since we are dealing with a group, there exist a number of sequences of 
transformations that can be used in principle to transform Minkowski space 
into the final SAV metric.  Certain transformations may be found to be easier 
to carry out than others, so, here, as in many things, experience is the best 
teacher.  However, if one wants a standardized treatment method, that would 
be obtained by employing an HKX transformation\cite{HKX}, either carried out 
as described by the authors of the HKX transformation, or by transforming 
Minkowski space into a static Weyl metric, a step that requires only a 
quadrature, then transforming that Weyl metric into the final SAV, a step 
that involves solving an ordinary Fredholm equation of the second 
kind\cite{he3}.  Moreover, a possible alternative approach has been provided 
by Klein and Richter\cite{KR}.

Even if the $\Pi$-matrix cannot be evaluated in closed form, by studying 
its domain of holomorphy in the extended $\tau$-plane one can ascertain 
whether or not an asymptotically flat singularity free global solution exists.  
In Sec.\ \ref{examples} we shall illustrate the various steps of the 
$\Pi$-matrix method with two examples.  The Schwarzschild interior solution
is ridiculously simple but has the advantage of involving only short 
and easy calculations.  The second example involves a rigidly rotating 
stationary axisymmetric dust spacetime with a dust density distribution
that is specified a priori.  There, since the pressure vanishes identically,
one has the option of selecting at will the matching surface.  After
choosing a simple surface, we were able to show that the matching SAV 
metric is certainly not asymptotically flat.  It would be exciting if 
one were to find a dust density distribution and a choice of matching 
surface that resulted in an asymptotically flat matching SAV metric.

Constructing a matching vacuum metric for the Wahlquist interior solution 
is likely to be challenging.  In that case we do not expect the matching 
SAV spacetime to be asymptotically flat.  We hope that the availability
of the general procedure for matching a given fluid solution to a vacuum 
metric that we shall describe in this paper will encourage those who 
discover a new exact SAPF solution to investigate possible physical 
interpretations of their fluid solution, even if it turns out that the 
matching SAV metric is not asymptotically flat.

\setcounter{equation}{0}
\section{The Perfect Fluid Field Equations\label{fieldeqs}} 

In this section we shall formulate the field equations that will be
assumed to hold both within the region occupied by a perfect fluid and 
within the vacuum region outside the source, and we shall stipulate our
continuity-differentiability premises.  

Let
\begin{equation} 
h_{cd} := \bK_{c}\cdot\bK_{d} \; (c,d \in \{3,4\}),
\label{2.4}
\end{equation}
where $\bK_{3}$ and $\bK_{4}$ denote the usual rotational and timelike 
Killing vector fields, respectively, whereupon $h_{33} > 0$ except on 
the axis, and $h_{44} < 0$ everywhere.  
Let $(\M^{2},g^{(2)})$ denote any one of the usual two dimensional
Riemannian subspaces\footnote{These are the surfaces of constant 
$(\varphi,t)$.} of the spacetime such that its Hausdorff space
$\M^{2}$ is orthogonal to the Killing vector orbits, and its atlas is
$\bC^{\infty}$.
All the differential forms that we employ in this paper will be on a two 
dimensional Hausdorff space $\M^{2}$ or on subspaces thereof.  In particular, 
$f$, $\omega$ and $P$ will denote the $0$-forms with domain $\M^{2}$ and values
\begin{equation} 
f(\x) = - h_{44}(\x), \; \omega(\x) := h_{34}(\x)/f(\x)
\end{equation}
and
\begin{equation} 
P(\x) := [h_{34}(\x)^{2} - h_{33}(\x) h_{44}(\x)]^{1/2} \ge 0
\label{2.13}
\end{equation}
for all $\x \in \M^{2}$.  Furthermore, $e_{12}$ will denote our
choice of the unit $2$-form on $\M^{2}$, and $\star$ will denote
the duality operator for the differential forms on $\M^{2}$ such
that
\begin{equation} 
\star e_{12} = 1, \; \star 1 = e_{12}
\end{equation}
and, for all $1$-forms $v$ and $w$ on $\M^{2}$,
\begin{equation} 
(\star v)w = e_{12} (v \cdot w),
\end{equation}
where $1$ denotes the identity mapping on $\M^{2}$, and we follow
the practice of suppressing `$\wedge$' in all exterior products and
derivatives of differential forms.  One readily shows that, for all
$1$-forms $v$ and $w$ on $\M^{2}$,
\begin{equation} 
\star \star v = - v, \; v(\star w) = w(\star v).
\end{equation}
Note that\footnote{This excludes event horizons from consideration.}
\begin{equation} 
P(\x) = 0 \text{ if and only if } \x \in \M^{1}_{ax},
\label{2.19}
\end{equation}
where $\M^{1}_{ax}$ denotes the axis subspace of $\M^{2}$.

\subsection{The Matching Premises}

In this first exposition of the $\Pi$-matrix formalism, we shall restrict
attention to those $\M^{2}$, the topology of which is described by Fig.\ 
1.\footnote{This restriction rules out, for example, sources of a toroidal 
shape.}  We assume that the common boundary $\M^{1}_{0}$ of the fluid interior 
$\M^{2}_{in}$ and vacuum exterior $\M^{2}_{ex}$ has a $\bC^{1}$ parametric 
equation
\begin{equation} 
\x = \x(\lambda) \; (0 \le \lambda \le \pi)
\end{equation}
that defines a homeomorphism of $[0,\pi]$ onto $\M^{1}_{0}$ such that the 
``north'' and ``south'' poles, where the zero pressure surface $\M^{1}_{0}$ 
and the axis $\M^{1}_{ax}$ intersect, are given by
\begin{equation} 
\bn = \x(0), \; \bs = \x(\pi).
\end{equation}
We also adopt the following (often tacitly assumed) premises:
\begin{equation}
\begin{array}{l}
\text{For all $\bC^{n}$ $1$-forms $v$ and $w$ on $\M^{2}$, $\star v$ and 
$v \cdot w$ are $\bC^{1}$ if $n \ge 1$ } \\
\text{and $\bC^{0}$ if $n = 0$.  Also, $p$ is $\bC^{0}$, $f$ and $\omega$ 
are $\bC^{1}$, while $P$ is $\bC^{2}$ } \\
\text{and $dP(\x) \ne 0$ at all $\x \in \M^{2}$.  Finally, there exists a 
continuous} \\
\text{$0$-form $\Lambda$ such that $P\Lambda$ is $\bC^{1}$ and 
$\omega = P^{2} \Lambda$.}
\end{array}
\label{premise1}
\end{equation}
These premises include all of the matching conditions at the zero pressure
surface $\M^{1}_{0}$, and they are consistent with the field equations that
we shall now specify.

\subsection{Four Field Equations and the Euler Equation}

It is convenient to introduce the $0$-forms 
\begin{equation} 
\Gamma := - \frac{1}{2} \ln (dP \cdot dP) \text{ and }
\psi := \frac{1}{2} \ln f
\label{2.24}
\end{equation}
and the $1$-forms
\begin{equation} 
\xi := - P^{-1} f^{2} \star d\omega 
:= - f^{2} \star [\Lambda dP + d(P \Lambda)]
\label{2.25}
\end{equation}
and
\begin{equation} 
\eta := P \star d\psi - \frac{1}{2} \omega \xi.
\label{2.26}
\end{equation}
In view of our premises, $\Gamma$ and $\psi$ are $\bC^{1}$, and $\xi$
and $\eta$ are $\bC^{0}$.  From the definition (\ref{2.24}) of $\Gamma$,
\begin{equation} 
e_{1} := e^{\Gamma} \star dP \text{ and } e_{2} := e^{\Gamma} dP
\end{equation}
constitute an orthonormal pair of $1$-forms on $\M^{2}$, and 
\begin{equation} 
e_{12} = e_{1}e_{2} = e^{2\Gamma} (\star dP) dP.
\end{equation}
We shall also employ the $0$-forms $\xi_{I}$, $\xi_{II}$, $d_{I}\psi$ 
and $d_{II}\psi$ for which 
\begin{equation}
\xi = (\xi_{I}) \star dP + (\xi_{II}) \; dP \text{ and }
d\psi = (d_{I}\psi) \star dP + (d_{II}\psi) \; dP.
\end{equation}

The restrictions to the fluid interior $\M^{2}_{in}$ and the vacuum exterior
$\M^{2}_{ex}$ of each of the differential forms on $\M^{2}$ that we have 
defined above will be denoted by the letter employed for the differential 
form with the affixed subscripts `$in$' and `$ex$', respectively.

We now introduce a function $\alpha$ that will be called the {\em boost
form} on $\M^{2}_{in}$.  Let $\bU$ denote the fluid world velocity field
on $\M^{4}_{in}$, where we recall that $\bU \cdot \bU = -1$.  Another
timelike unit vector on $\M^{4}_{in}$ is the restriction to $\M^{4}_{in}$
of
\begin{equation} 
\be_{t} := - \frac{1}{\sqrt{f}} \bK_{4}.
\end{equation}
It is easy to show that $|\be_{t} \cdot \bU| \ge 1$ throughout $\M^{4}_{in}$.
We select the sign of $\bK_{4}$ so that $\be_{t} \cdot \bU \ge 1$ and
let $\alpha$ denote the $0$-form on $\M^{2}_{in}$ such that
\begin{equation} 
\cosh{\alpha(\x)} := \be_{t} \cdot \bU(\x), \quad
\sgn{\alpha} := \sgn{U^{3}}
\end{equation}
for all $\x \in \M^{2}_{in}$.  The boost form will be employed in our
expressions for the field equations.

To ensure that the field equations are meaningful at all points of
$\M^{2}_{in} \cup \M^{2}_{ex}$ including points on $\M^{1}_{ax}$, we adopt 
the following premises, which are not necessarily independent of one another:
\begin{equation}
\begin{array}{l}
\text{For all $\bC^{\infty}$ $1$-forms $v$ and $w$ on $\M^{2}_{in} \cup 
\M^{2}_{ex}$, $\star v$ and $v \cdot w$ are $\bC^{\infty}$.} \\
\text{Also, $f_{in}$, $f_{ex}$, $P_{in}$, $P_{ex}$, $\Lambda_{in}$, 
$\Lambda_{ex}$, $p_{in}$, $\epsilon_{in}$ and $P_{in}^{-1} \sinh{\alpha}$ are 
$\bC^{\infty}$.}
\end{array}
\label{premise2}
\end{equation}
These premises are certainly stronger than necessary, but they enable
us to avoid complications that would obscure this first exposition of
the $\Pi$-matrix method.

With the above premises, all but one of the field equations are as 
follows:
\begin{eqnarray} 
d(\Gamma+\psi) - P \left\{ (\star d\psi) d_{I}\psi
+ (d\psi) d_{II}\psi \right. & & \nonumber \\
\left. + (2f)^{-2} [(\star \xi) \xi_{I} + (\xi) \xi_{II}]
\right\} & = & - \kappa P p e^{2\Gamma} dP,
\label{2.32}
\end{eqnarray}
\begin{eqnarray} 
d \star dP & = & - 2 \kappa P p e_{12},
\label{2.33} \\
d\xi & = & \kappa f (p + \epsilon) (\sinh{2\alpha}) e_{12},
\label{2.34} \\
d\eta & = & - \frac{1}{2} \kappa P [2p + (p + \epsilon) \cosh{2\alpha}]
e_{12},
\label{2.35}
\end{eqnarray}
where $\kappa := 8\pi G/c^{4}$ in conventional metric units, and it is
to be understood that the restriction of the right sides of Eqs.\ (\ref{2.34})
and (\ref{2.35}) to $\M^{2}_{ex}$ are identically zero.  The contracted Bianchi
identity in $\M^{4}_{ex}$ is identically satisfied, and in $\M^{4}_{in}$
it is equivalent to the following general relativistic Euler equation
(in which we suppress the subscripts `$in$'): 
\begin{equation} 
dp + (p + \epsilon) [(\cosh{2\alpha})d\psi - (\cosh{2\alpha}-1)dP/2P
- (\sinh{2\alpha})\star\xi/2f] = 0.
\label{2.36}
\end{equation}

Note that the field equations (\ref{2.32}) and (\ref{2.33}) hold throughout 
$\M^{2}$ including $\M^{1}_{0}$.  From Eqs.\ (\ref{2.19}) and  (\ref{2.32}), 
$\Gamma(\x) + \psi(\x) = \Gamma(\bn) + \psi(\bn)$ at all $\x \in \M^{1}_{ax}$.
As is well known, the condition
\begin{equation} 
\Gamma(\x) + \psi(\x) = 0 \text{ at all } \x \in \M^{1}_{ax}
\end{equation}
is\footnote{granted our premises} necessary and sufficient for $g$ to be 
locally Minkowskian on $\M^{2}_{ax}$; i.e., to have no ``conical 
singularities'' on $\M^{2}_{ax}$.  Therefore, we shall henceforth assume 
that $\Gamma(\bn) + \psi(\bn) = 0$.

Unlike Eqs.\ (\ref{2.32}) and (\ref{2.33}), the field equations 
(\ref{2.34}) and (\ref{2.35}) are generally defined only in the domain 
$\M^{2}_{in} \cup \M^{2}_{ex}$.  This situation changes when $\epsilon(\x) = 0$ 
at all $\x \in \M^{1}_{0}$, but we shall not pursue that topic in this paper.

It is desirable to extend the domains of all of the field equations in
$\M_{in}^{2}$ to $\bar{\M}_{in}^{2} := \M_{in}^{2} \cup \M_{0}^{1}$,
of all of the field equations in
$\M_{ex}^{2}$ to $\bar{\M}_{ex}^{2} := \M_{ex}^{2} \cup \M_{0}^{1}$,
and of the Euler equation (\ref{2.36}) to $\bar{\M}_{in}^{2}$.  To
accomplish this, it is sufficient\footnote{It is to be understood
that all derivatives at boundary points of $\bar{\M}_{in}^{2}$ and of
$\bar{\M}_{ex}^{2}$ are defined using only sequences of points in
$\bar{\M}_{in}^{2}$ and $\bar{\M}_{ex}^{2}$, respectively.} to introduce
the following reasonable premises, which are consistent with all other
premises of this paper and with Eqs.\ (\ref{2.32}) to (\ref{2.36}):
\begin{equation}
\begin{array}{l}
\text{For all $\bC^{2}$ $1$-forms $v$ and $w$ on $\bar{\M}_{in}^{2}$,
$\star v$ and $v \cdot w$} \\
\text{are $\bC^{2}$; and $P_{in}$, $f_{in}$,
$\Lambda_{in}P_{in}$, $\Lambda_{in}$, $p_{in}$, $\epsilon_{in}$ and} \\
\text{$P_{in}^{-1} \sinh{\alpha}$ have $\bC^{3}$, $\bC^{2}$, $\bC^{2}$,
$\bC^{1}$, $\bC^{1}$, $\bC^{0}$ and $\bC^{0}$} \\
\text{extensions, respectively, to $\bar{\M}_{in}^{2}$.  For all 
$\bC^{\infty}$ $1$-forms} \\
\text{$v$ and $w$ on $\bar{\M}_{ex}^{2}$, $\star v$ and $v \cdot w$ 
are $\bC^{\infty}$; and $P_{ex}$, $f_{ex}$} \\
\text{and $\Lambda_{ex}$ have $\bC^{\infty}$ extensions to 
$\bar{\M}_{ex}^{2}$.}
\end{array}
\label{premise3}
\end{equation}

The only remaining field equation (one which we have chosen to suppress
in this paper) happens to be a $2$-form equation involving $d \star d\Gamma$.
A well known proposition asserts that, if $(f_{ex},P_{ex},\Lambda_{ex})$
is a solution of Eqs.\ (\ref{2.33}), (\ref{2.34}) and (\ref{2.35}) on
$\bar{\M}_{ex}^{2}$, then\footnote{Note that we are taking the liberty of 
employing the same notations for differential forms on $\bar{\M}_{in}^{2}$ 
and $\bar{\M}_{ex}^{2}$ as we do for their restrictions to $\M_{in}^{2}$ and 
$\M_{ex}^{2}$, respectively.  However, keep in mind that the differential 
forms that we are considering are $\bC^{\infty}$ in $\M_{in}^{2} \cup 
\M_{ex}^{2}$, but are not necessarily $\bC^{\infty}$ on $\M_{0}^{1}$.} 
Eq.\ (\ref{2.32}) on $\bar{\M}_{ex}^{2}$ is completely integrable; and,
if $\Gamma_{ex}$ is the integral of Eq.\ (\ref{2.32}) on $\bar{\M}_{ex}^{2}$,
then $(f_{ex},P_{ex},\Lambda_{ex},\Gamma_{ex})$ identically satisfies the
$d \star d\Gamma$ field equation on $\bar{\M}_{ex}^{2}$.  It can also be 
shown that if 
$(f_{in},P_{in},\Lambda_{in},\Gamma_{in},p_{in},\epsilon_{in},\alpha)$ 
is a solution of Eqs.\ (\ref{2.33}), (\ref{2.34}), (\ref{2.35}), (\ref{2.36}) 
and (suppressing the subscript `in')
\begin{equation}
d_{I}(\Gamma+\psi) - 2 P \left\{ d_{I}\psi \; d_{II}\psi + (2f)^{-2}
\xi_{I} \; \xi_{II} \right\} = 0 
\end{equation}
on $\bar{\M}_{in}^{2}$, then\footnote{Ibid} 
Eq.\ (\ref{2.32}) on $\bar{\M}_{in}^{2}$ is completely integrable, and, if 
$\Gamma_{in}$ is the integral of Eq.\ (\ref{2.32}) on $\bar{\M}_{in}^{2}$, 
then $(f_{in},P_{in},\Lambda_{in},\Gamma_{in},p_{in},\epsilon_{in})$ 
identically satisfies the $d \star d\Gamma$ field equation on 
$\bar{\M}_{in}^{2}$.

\subsection{The Vacuum Region}

We now focus attention on Eqs.\ (\ref{2.33}), (\ref{2.34}) and (\ref{2.35}) 
on $\bar{\M}^{2}_{ex}$.  Since $\bar{\M}^{2}_{ex}$ is simply connected, 
these field equations are equivalent to the statement\footnote{We are taking 
the liberty of suppressing the subscripts `$ex$' in `$P_{ex}$', `$\xi_{ex}$' 
and `$\eta_{ex}$'.  Similar abbreviating liberties will be employed in later 
equations, and we shall depend on the context to help the reader avoid 
confusion.} that $\bC^{\infty}$ $0$-forms $Z$, $\chi$ and $\phi$ 
exist on $\bar{\M}^{2}_{ex}$ such that
\begin{eqnarray} 
\star dP & = & dZ, 
\label{2.38} \\
\xi & = & d\chi, 
\label{2.39} \\
\eta & = & d\phi,
\label{2.40}
\end{eqnarray}
Equation (\ref{2.38}) enables us to introduce the Weyl canonical chart
$\x \rightarrow (z,\rho) := (Z(\x),P(\x))$ that maps $\bar{\M}_{ex}^{2}$
onto
\begin{equation}
\bar{\D}_{ex} := \{(Z(\x),P(\x)): \x \in \bar{\M}_{ex}^{2}\}.
\label{2.DEX}
\end{equation}
As is well known, Eqs.\ (\ref{2.39}) and (\ref{2.40}) then yield the
following elliptic Ernst equation for the potential $\E := f + i \chi$
expressed as a function $\E(z,\rho) = f(z,\rho) + i \chi(z,\rho)$ of
Weyl canonical coordinates:
\begin{eqnarray} 
f(z,\rho) \left\{ \frac{\partial}{\partial z} 
\left[ \rho \frac{\partial\E(z,\rho)}{\partial z}\right]
+ \frac{\partial}{\partial\rho} 
\left[ \rho \frac{\partial\E(z,\rho)}{\partial\rho}\right]\right\}
\nonumber \\
\mbox{ } + \rho \left\{ \left(\frac{\partial\E(z,\rho)}{\partial z}\right)^{2}
+ \left(\frac{\partial\E(z,\rho)}{\partial\rho}\right)^{2}\right\} & = & 0
\label{2.47}
\end{eqnarray}
throughout $\bar{\D}_{ex}$.  We shall describe a way to determine the
axis values of this $\E$-potential from the matching data at the zero
pressure surface.  Many methods are known that permit the construction
of $\E(z,\rho)$ from $\E(z,0)$, assuming that suitable premises are
satisfied.  It is, however, not our purpose to rehash these well known 
procedures.

\setcounter{equation}{0}
\section{The Monodromy Matrix $\Pi(\tau)$\label{mat}} 
\subsection{Determination of the Matching Data} 

The curve 
\begin{equation}
\D_{0} := \{(Z(\x),P(\x)):\x \in \M^{1}_{0}\}
\label{SP0}
\end{equation}
that represents the matching surface is given by the parametric equations
\begin{equation} 
z(\lambda) := Z(\x(\lambda)), \; \rho(\lambda) := P(\x(\lambda))
\quad (0 \le \lambda \le \pi),
\label{2.43}
\end{equation}
and the matching data comprise $z(\lambda)$, $\rho(\lambda)$ and
\begin{equation} 
\begin{array}{l}
f(\lambda) := f(\x(\lambda)), \; \chi(\lambda) := \chi(\x(\lambda)), \\
\omega(\lambda) := \omega(\x(\lambda)), \; \phi(\lambda) := \phi(\x(\lambda)).
\end{array}
\label{2.44}
\end{equation}
The functions of $\lambda$ that are defined above are to be determined
from the given fluid solution in $\bar{\M}^{2}_{in}$ as follows.

The given fluid solution is a compact two-dimensional Riemannian space
$(\bar{\M}^{2},g_{in}^{(2)})$ with a maximal $\bC^{\infty}$ atlas.  If
$(x^{1},x^{2})$ denotes the coordinate pair corresponding to any point $\x$
in the domain of a chart in this atlas, one can use the metric $g_{in}^{(2)}$
to compute the $0$-forms $\Y_{j}^{i}$ for which 
\begin{equation}
\star dx^{i} = dx^{j} \Y_{j}^{i}(\x).
\end{equation}
Then we may write
\begin{equation}
(\star dP)(\x) = dx^{j} \Y_{j}^{i}(\x)P_{i}(\x),
\end{equation}
where $P_{i}(\x)$ is defined by
\begin{equation}
dP(\x) = dx^{i} P_{i}(\x).
\end{equation}
The pull-back of $\star dP$ corresponding to the mapping $\lambda \rightarrow
\x(\lambda)$ of $[0,\pi]$ into $\bar{\M}_{in}^{2}$ is, therefore, given by
\begin{equation}
dz(\lambda) := d\lambda \dot{x}^{i}(\lambda) \Y_{i}^{j}(\x(\lambda))
P_{i}(\x(\lambda))
\end{equation}
for each chart whose domain contains an interval of $\M_{0}^{1}$, where 
$x^{1}(\lambda)$ and $x^{2}(\lambda)$ denote the coordinates of the point
$\x(\lambda)$ in the interval.  Integration of $dz(\lambda)$ now yields
$z(\lambda)$ up to an additive constant $z(0)$ whose value can be chosen
freely to fit some convention.  On the other hand,
\begin{equation}
\rho(\lambda) := P(\x(\lambda)).
\end{equation}
The pairs $(\chi(\lambda),\omega(\lambda))$ and $(\phi(\lambda),f(\lambda))$
are treated similarly to how we have here treated the pair 
$(z(\lambda),\rho(\lambda))$.  By specifying all six of these objects, we
are, in effect, specifying both the tangential and normal derivatives of
$\rho(\x)$, $f(\x)$ and $\omega(\x)$ on the zero pressure surface.
Moreover, nothing is assumed concerning behavior of the solution at large 
distances from the fluid source.  What we are solving then is neither a 
Dirichlet nor a Neumann problem.

\subsection{A Linear System for the Ernst Equation}

In a recent paper\cite{recent} we described, and showed the relationships
among, three linear systems for the vacuum Ernst equation, one inferred 
by the present authors from the Kinnersley-Chitre formalism, one developed 
and used extensively by Neugebauer and a new one that the present authors 
have found valuable in the course of developing formal proofs, especially 
in connection with the hyperbolic Ernst equation.\footnote{In the present 
paper our $\Delta(\x,\tau)$ and $\Q(\x,\tau)$ are those fields that were 
denoted by $\Gamma_{HE}(\x,\tau)$ and $\F_{HE}(\x,\tau)$, respectively, in 
the cited reference.}  It is the last named one that we plan also to use 
here.  It will be expressed in the form
\begin{equation}
d\Q(\x,\tau) = \Delta(\x,\tau) \Q(\x,\tau)
\label{linsys}
\end{equation}
for all $\x \in \bar{\M}_{ex}^{2}$ and $\tau \in C - \bar{K}(\x)$,
where it is understood that $d$ does not operate on the complex spectral
parameter $\tau$ (i.e., $d\tau = 0$) and where
\begin{eqnarray}
\Delta(\x,\tau) & := & -\left(\frac{\tau-Z(\x)+P(\x)\star}{\mu(\x,\tau)}
\right) \left(\frac{I df(\x) - J d\chi(\x)}{2f(\x)}\right) \sigma_{3} 
\nonumber \\ & & \mbox{ } - J \frac{d\chi(\x)}{2f(\x)},
\label{Del}
\end{eqnarray}
where
\begin{equation}
I = \left( \begin{array}{cc}
1 & 0 \\ 0 & 1
\end{array} \right), \quad
J = \left( \begin{array}{cc}
0 & 1 \\ -1 & 0
\end{array} \right), \quad
\sigma_{3} = \left( \begin{array}{cc}
1 & 0 \\ 0 & -1
\end{array} \right),
\end{equation}
\begin{equation}
\mu(\x,\tau) := \left[(\tau-Z(\x))^{2} + P(\x)^{2}\right]^{1/2}, \quad
\lim_{\tau\rightarrow\infty} \frac{\mu(\x,\tau)}{\tau} = 1,
\end{equation}
and the cut $\bar{K}(\x)$ in the complex $\tau$-plane is a simple
$\bC^{1}$ arc whose end points are the branch points $Z(\x) \pm i P(\x)$
of $\mu(\x,\tau)$.  Moreover, $\bar{K}(\x)$ is symmetric with respect to
the real axis, is a subset of\footnote{See Eq.\ (\ref{2.DEX}).}
\begin{equation}
\bar{\Sigma}_{ex} := \{z \pm i\rho:(z,\rho) \in \bar{\D}_{ex}\}
\label{3.SIGEX}
\end{equation}
and intercepts the real axis at a point on the same side of the closed
contour
\begin{equation}
\Sigma_{0} := \{z(\lambda) \pm i \rho(\lambda):0 \le \lambda \le \pi\}
\label{3.SIG0}
\end{equation}
as the point $z(0)$ that represents the north pole $\bn$, i.e., the
point of interception is in
\begin{equation}
\Sigma_{ax+} := \{z \in R^{1}:(z,0) \in \bar{\D}_{ex} \text{ and }
z \ge z(0)\}.
\label{3.AX}
\end{equation}

The solution of Eq.\ (\ref{linsys}) will be made unique by specifying
that 
\begin{equation}
\Q(\bn,\tau) := e^{-\sigma_{3}\psi(\bn)} \left( \begin{array}{cc}
1 & - \chi(\bn) \\ 0 & 1
\end{array} \right)
\label{3.QNP}
\end{equation}
for all $\tau \in C$.  We reserve the option of scaling $x^{4}=t$ 
and choosing the arbitrary additive real constant in $\chi$ so that
$\E(\bn) = 1$, whereupon $\Q(\bn,\tau)=I$.
To grasp the motivation behind the selection (\ref{3.QNP}), note
that the Ernst equation (\ref{2.47}) implies that 
$\partial\E(z,\rho)/\partial\rho$ vanishes at $\rho = 0$.  Therefore,
if one expresses all differential forms in Eqs.\ (\ref{linsys}) and
(\ref{Del}) as functions of $(z,\rho,\tau)$,
\begin{equation}
\Delta(z,0,\tau) = - \sigma_{3} d\psi(z,0) - \left( \begin{array}{cc}
0 & 1 \\ 0 & 0 
\end{array} \right) e^{-2\psi(z,0)} d\chi(z,0).
\label{3.1AX}
\end{equation}
Therefore, the solution on $\{(z,0) \in \bar{\D}_{ex}:z \ge z(0)\}$
of Eqs.\ (\ref{linsys}) and (\ref{3.QNP}) is
\begin{equation}
\Q(z,0,\tau) = e^{-\sigma_{3}\psi(z,0)} \left( \begin{array}{cc}
1 & -\chi(z,0) \\ 0 & 1
\end{array} \right) \text{ for all $z \in \Sigma_{ax+}$ and $\tau
\in C$.}
\label{3.2AX}
\end{equation}

Three key properties of $\Q(\x,\tau)$ are easily deducible from Eqs.\ 
(\ref{linsys}) and (\ref{3.QNP}).  First, since $\Q(\bn,\tau)$ is real
and $\Delta(\x,\tau) := \Delta(\x,\tau^{*})^{*} = \Delta(\x,\tau)$,
the reality condition
\begin{equation}
\Q^{*}(\x,\tau) := [\Q(\x,\tau^{*})]^{*} = \Q(\x.\tau)
\label{3.REAL}
\end{equation}
holds.  Second, since $\tr{\Delta(\x,\tau)}=0$ and $\det{\Q(\bn,\tau)}=1$,
\begin{equation}
\det{\Q(\x,\tau)}=1.
\label{3.DET}
\end{equation}
Third, for fixed $\x \in \bar{\M}_{ex}^{2}$, $\Q(\x,\tau)$ is a 
holomorphic function of $\tau$ throughout $C - \bar{K}(\x)$.

Employing Eqs.\ (\ref{2.25}), (\ref{2.26}), (\ref{2.39}), 
(\ref{2.40}) and (\ref{Del}), one can express $\Delta(\x,\tau)$
in the following form that no longer contains the duality operator:
\begin{eqnarray}
\Delta(\x,\tau) & := & -\frac{\tau-Z(\x)}{\mu(\x,\tau)}
\left(\frac{I df(\x) - J d\chi(\x)}{2f(\x)}\right) \sigma_{3} 
\nonumber \\ & & \mbox{ } - \frac{1}{\mu(\x,\tau)} 
\left[I \left(d\phi(\x) + \frac{1}{2}\omega(\x) d\chi(\x)\right) 
- \frac{1}{2} J f(\x) d\omega(\x)\right] \sigma_{3}
\nonumber \\ & & \mbox{ } - J \frac{d\chi(\x)}{2f(\x)}.
\label{Delta}
\end{eqnarray}
Thus, we conclude that
\begin{eqnarray}
\frac{\partial \Q(\x(\lambda),\tau)}{\partial\lambda} & = & 
\left\{-\frac{\tau-z(\lambda)}{\mu(\x(\lambda),\tau)}
\left(I \frac{\dot{f}(\lambda)}{2f(\lambda)}
- J \frac{\dot{\chi}(\lambda)}{2f(\lambda)}\right)
\right.  \sigma_{3} \nonumber \\ 
& & \mbox{ } - \frac{1}{\mu(\x(\lambda),\tau)} 
\left[I \left(\dot{\phi}(\lambda)
+ \frac{1}{2}\omega(\lambda) \dot{\chi}(\lambda)\right) \right. \nonumber \\
& & \left. \mbox{ } - \frac{1}{2} J f(\lambda) \dot{\omega}(\lambda)
\right] \sigma_{3} 
\nonumber \\ & & \mbox{ } \left.
- J \frac{\dot{\chi}(\lambda)}{2f(\lambda)} \right\} 
\Q(\x(\lambda),\tau) \text{ for all } 0 \le \lambda \le \pi \nonumber \\
& & \text{ and } \tau \in C - \bar{K}(\x(\lambda)),
\label{diffeq}
\end{eqnarray}
where $\Q(\x(0),\tau) = \Q(\bn,\tau)$ is given by Eq.\ (\ref{3.QNP}),
where dots denote derivatives with respect to $\lambda$, and where
$\bar{K}(\x(\lambda))$ lies on $\Sigma_{0}$ and is
\begin{equation}
\bar{K}(\x(\lambda)) = \{z(\lambda') \pm i \rho(\lambda'):0 \le \lambda'
\le \lambda\}.
\end{equation}
We note that only the matching data (\ref{2.43}) and (\ref{2.44})
are needed in order to be able to write out this ordinary differential
equation.  We shall employ the convenient abbreviation
\begin{equation}
\Q(\lambda,\tau) := \Q(\x(\lambda),\tau).
\label{3.q}
\end{equation}

Thus, both the closed contour $\Sigma_{0}$ and the function $\Q(\lambda,\tau)$
are uniquely determined by the matching data $z(\lambda)$, $\rho(\lambda)$,
$\chi(\lambda)$, $\omega(\lambda)$, $\phi(\lambda)$ and $f(\lambda)$, the
differential equation (\ref{diffeq}) and the initial condition
\begin{equation}
\Q_{0} := \Q(0,\tau) := e^{\sigma_{3}\psi(0)} \left( \begin{array}{cc}
1 & - \chi(0) \\ 0 & 1
\end{array} \right).
\label{3.qIC}
\end{equation}

\subsection{Determination of the $\Pi$-Matrix}

It is from the function $\Q(\lambda,\tau)$ that one computes the
monodromy matrix $\Pi(\tau)$ provided that the latter exists.  A class
of spinning fluid spheroids for which monodromy matrices exist will now
be defined.  Specifically, we shall henceforth consider the class of all
SAPF solutions $(\bar{\M}_{in}^{2},g_{in}^{(2)})$ for which the following
two sets of conditions hold:
\begin{romanlist}
\item
The parts of the premises (\ref{premise2}) and (\ref{premise3}) that 
concern $\M_{in}^{2}$ and $\bar{\M}_{in}^{2}$, respectively, are satisfied.
Also, $\dot{z}(\lambda)^{2} + \dot{\rho}(\lambda)^{2} > 0$ for all
$\lambda \in [0,\pi]$.
\item
As regards the function $\Q$, there exist $2 \times 2$ matrix
functions $\Q_{1}$ and $\Q_{2}$ with a common domain $[0,\pi] \times 
\hat{\Sigma}$ such that $\hat{\Sigma} \subset C$ and is an open covering
of $\Sigma_{0}$ that is either topologically equivalent to an annulus or
is a simply connected open neighborhood of $\infty$, $\hat{\Sigma}^{*}
= \hat{\Sigma}$,
\begin{equation}
\begin{array}{l}
\Q(\lambda,\tau) = \Q_{1}(\lambda,\tau) + \mu(\x(\lambda),\tau) 
\Q_{2}(\lambda,\tau) \\
\text{for all } \lambda \in [0,\pi] \text{ and }
\tau \in \hat{\Sigma} - \bar{K}(\x(\lambda))
\end{array}
\label{3.q12}
\end{equation}
and, for each $\tau \in \hat{\Sigma}$, $\Q_{1}(\lambda,\tau)$ and
$\Q_{2}(\lambda,\tau)$ are $\bC^{1}$ functions of $\lambda$ throughout
$[0,\pi]$.  Also, for each $\lambda \in [0,\pi]$, $\Q_{i}(\lambda,\tau)$
and $\partial\Q_{i}(\lambda,\tau)/\partial\lambda$ $(i=1,2)$ are 
holomorphic functions of $\tau$ throughout $\hat{\Sigma}$.
\end{romanlist}

That completes the two sets of conditions.  Conditions (ii) are required
for the existence of the monodromy matrix.  To illustrate how the 
conditions (ii) are employed in proofs, consider the fact that
$\Q^{*}(\lambda,\tau) = \Q(\lambda,\tau)$ for all $\lambda \in [0,\pi]$
and $\tau \in C - \bar{K}(\x(\lambda))$.\footnote{The derivation of this
fact is similar to that of Eq.\ (\ref{3.REAL}).}  Therefore, from Eq.\ 
(\ref{3.q12}),
\begin{equation}
\begin{array}{l}
\left[\Q_{1}^{*}(\lambda,\tau)-\Q_{1}(\lambda,\tau)\right] + \mu(\x(\lambda,\tau)
\left[\Q_{2}^{*}(\lambda,\tau)-\Q_{2}(\lambda,\tau)\right] = 0 \\
\text{for all } \lambda \in [0,\pi] \text{ and }
\tau \in \hat{\Sigma} - \bar{K}(\x(\lambda)).
\end{array}
\label{3.REAL.1}
\end{equation}
For each $0 < \lambda < \pi$, analytic continuation on any simple closed
orbit that lies in $\hat{\Sigma}$ and that encloses one and only one of
the branch points $z(\lambda) \pm i \rho(\lambda)$ of $\mu(\x(\lambda),\tau)$
exists and induces the replacements $\Q_{i}(\lambda,\tau) \rightarrow
\Q_{i}(\lambda,\tau)$ and $\mu(\x(\lambda),\tau) \rightarrow 
-\mu(\x(\lambda),\tau)$ in Eq.\ (\ref{3.REAL.1}), whereupon one obtains
$\Q_{i}^{*}(\lambda,\tau) = \Q_{i}(\lambda,\tau)$ for all $i \in \{1,2\}$,
$0 < \lambda < \pi$ and $\tau \in \hat{\Sigma} - \bar{K}(\x(\lambda))$.
The continuity properties of $\Q_{i}$ then yields
\begin{equation}
\Q_{i}^{*}(\lambda,\tau) = \Q_{i}(\lambda,\tau) \text{ for all } 
i \in \{1,2\}, \lambda \in [0,\pi] \text{ and } \tau \in \hat{\Sigma}.
\label{3.REAL.2}
\end{equation}

For each $0 < \lambda < \pi$, analytic continuation of (\ref{3.q12}) on
any simple closed orbit in $\hat{\Sigma}$ that encloses exactly one of
the points $z(\lambda) \pm i \rho(\lambda)$ is seen to exist and yields
\begin{equation}
\Q'(\lambda,\tau) := \Q_{1}(\lambda,\tau) - \mu(\x(\lambda),\tau)
\Q_{2}(\lambda,\tau).
\label{3.qAC}
\end{equation}
Then, by letting $\Q'_{i}(0,\tau)$ and $\Q'_{i}(\pi,\tau)$ denote the
limits of $\Q'(\lambda,\tau)$ as $\lambda \rightarrow 0$ and $\lambda
\rightarrow \pi$, respectively, we extend Eq.\ (\ref{3.qAC}) to all
$\lambda \in [0,\pi]$ and $\tau \in \bar{K}(\x(\lambda))$.  
Since\footnote{See Eq.\ (\ref{3.DET}).} $\det{\Q(\lambda,\tau)} = 1$,
it follows by employing analytic continuation and the continuity properties 
of $\Q_{i}$ in the manner illustrated by the proof of Eq.\ (\ref{3.REAL.2})
that
\begin{equation}
\det{\Q'(\lambda,\tau)} = \det{\Q(\lambda,\tau)} = 1 
\text{ for all } \lambda \in [0,\pi] \text{ and }
\tau \in \hat{\Sigma} - \bar{K}(\x(\lambda)).
\label{3.qdet}
\end{equation}
Furthermore, by employing analytic continuation together with the
continuities as functions of $\lambda$ of $\Q_{i}(\lambda,\tau)$ and
$\partial \Q_{i}(\lambda,\tau)/\partial\lambda$, one proves that
$-J \Q'(\lambda,\tau) J$ satisfies the same differential equation
(\ref{diffeq}) as $\Q(\lambda,\tau)$ for all $\lambda \in [0,\pi]$ 
and $\tau \in \hat{\Sigma} - \bar{K}(\x(\lambda))$.  Thus, $\Pi(\tau)$
exists for each $\tau \in \hat{\Sigma}$ such that
\begin{equation}
-J \Q'(\lambda,\tau) J = \Q(\lambda,\tau) \Pi(\tau)
\label{3.PI}
\end{equation}
for all $\lambda \in [0,\pi]$ and $\tau \in \hat{\Sigma} - \bar{K}(\x(\lambda))$.
By then employing Eq.\ (\ref{3.qdet}) and the fact that $M^{T} J M =
(\det{M})J$ for any $2 \times 2$ matrix $M$, one obtains
\begin{equation}
\Pi(\tau) = - J \Q(\lambda,\tau)^{T} \Q'(\lambda,\tau) J.
\label{3.PI.1}
\end{equation}
Therefore, with the aid of Eqs.\ (\ref{3.q12}) and (\ref{3.qAC}), we obtain
the convenient formula
\begin{equation}
\Pi(\tau) = - J \Q_{0}^{T} \Q'(0,\tau) J = - J \Q_{0}^{T} [2\Q_{1}(0,\tau)-\Q_{0}],
\label{3.PI.2}
\end{equation}
from which one sees that $\Pi(\tau)$ is holomorphic throughout $\hat{\Sigma}$.

From Eqs.\ (\ref{3.qdet}) and (\ref{3.PI}) the determinantal property 
\begin{equation}
\det{\Pi(\tau)} = 1
\label{3.PIdet}
\end{equation}
follows, while from Eqs.\ (\ref{3.qIC}), (\ref{3.REAL.2}) and (\ref{3.PI.2})
the reality property 
\begin{equation}
\Pi^{*}(\tau) = \Pi(\tau)
\label{3.PIcc}
\end{equation}
follows.  Moreover, from Eq.\ (\ref{3.PI.1}) and the fact that analytic 
continuation on a simple closed orbit that encloses one and only one of the 
branch points $z(\lambda) \pm i \rho(\lambda)$ induces $\Q(\lambda,\tau) 
\rightarrow \Q'(\lambda,\tau)$ and $\Q'(\lambda,\tau) \rightarrow 
\Q(\lambda,\tau)$ the symmetry property 
\begin{equation}
\Pi(\tau)^{T} = \Pi(\tau)
\label{3.PIsymm}
\end{equation}
follows.

Upon letting $\lambda = 0$ and $\tau = z(0)$ in Eqs.\ (\ref{3.q12}) and then
using Eq.\ (\ref{3.qIC}), we obtain $\Q_{1}(0,z(0)) = \Q_{0}$.  Therefore,
from Eq.\ (\ref{3.PI.2}),
\begin{equation}
\Pi(z(0)) = - J \Q_{0}^{T} \Q_{0} J.
\label{3.PI.3}
\end{equation}
It is clear from Eqs.\ (\ref{3.PIdet}) to (\ref{3.PI.3}) that
\begin{equation}
\begin{array}{l}
\text{$\Pi(\tau)$ is unimodular, real, symmetric and } \\
\text{positive definite for all $\tau$ in the maximal } \\
\text{real subinterval of the real axis in $\hat{\Sigma}$ } \\
\text{that contains $z(0)$.}
\end{array}
\label{3.PI.4}
\end{equation}

\subsection{Use of $\Pi(\tau)$ to Assess Asymptotic Flatness}

It is unnecessary to construct the exterior vacuum solution in order 
to determine if that solution is asymptotically flat, for this can
be determined by investigating instead the analytic properties of 
$\Pi(\tau)$.  To help describe this investigation, we shall need the
following concepts: 
\begin{eqnarray}
\Sigma_{EXT} & := & \text{ that open unbounded subset of $C$} 
\nonumber \\
& & \text{ whose boundary is $\Sigma_{0}$,} \\
\bar{\Sigma}_{EXT} & := & \Sigma_{EXT} \cup \Sigma_{0}, \\
\bar{S}_{EXT} & := & \{(z,\rho) \in R^{2}: (z + i\rho) \in 
\bar{\Sigma}_{EXT}\}, \\
\bar{\D}_{EXT} & := & \{(z,\rho) \in \bar{S}_{EXT}: \rho \ge 0\}
\end{eqnarray}
and
\begin{eqnarray}
\bar{\Sigma}_{ex} & := & \bar{\Sigma}_{EXT} \cap \hat{\Sigma}, \\
\bar{S}_{ex} & := & \{(z,\rho) \in \bar{S}_{EXT}: z+i\rho \in 
\bar{\Sigma}_{ex}\}, \\
\bar{\D}_{ex} & := & \{(z,\rho) \in \bar{S}_{ex}: \rho \ge 0\}.
\end{eqnarray}

There are four distinct cases:
\begin{romanlist}
\item
Suppose that $\bar{\Sigma}_{ex} = \bar{\Sigma}_{EXT}-\{\infty\}$, but
$\Pi_{44}(\tau)$ and $\tau^{-1}\Pi_{34}(\tau)$ have holomorphic
extensions that cover $\tau=\infty$.  The transformation $\bK_{4}
\rightarrow \bK_{4}+k\bK_{3}$, where $k$ is a real number, induces
$\Pi_{44}(\tau) \rightarrow \Pi_{44}(\tau)$ and $\Pi_{34}(\tau)
\rightarrow \Pi_{34}(\tau) - 2k\tau$.  So, this transformation can
be used to make $\Pi(\tau)$ holomorphic at $\tau=\infty$, whereupon
$\bar{\Sigma}_{ex}=\bar{\Sigma}_{EXT}$.   Then we can and we do scale
$\bK_{4}$ and select $\chi(0)$ so that
\begin{equation}
\Pi(\infty) = I.
\end{equation}
If it is also true that the matrix elements $\Pi_{ab}(\tau)$ satisfy
\begin{equation}
\tau \Pi_{34}(\tau)/\Pi_{44}(\tau) \rightarrow 0 \text{ as }
\tau \rightarrow \infty,
\label{lim}
\end{equation}
then the Ernst potential $\E$ of the SAV metric expressed as a function of 
Weyl canonical coordinates can be analytically extended to a domain that 
covers $\bar{S}_{EXT}$ such that $\E(z,-\rho)=\E(z,\rho)$, and the 
restriction of $\E$ to $\bar{\D}_{EXT}$ yields a SAV metric without 
singularities that is asymptotically flat and satisfies all of the requisite 
matching conditions at the zero pressure surface.  So, in this case, we 
obtain an asymptotically flat global solution.

\item
Suppose that $\bar{\Sigma}_{ex} = \bar{\Sigma}_{EXT}$ as in the preceding
case, but the condition (\ref{lim}) does not hold.  Then, again, $\E$ has 
an analytic continuation to a domain that covers $\bar{S}_{EXT}$ such that 
$\E(z.-\rho) = \E(z,\rho)$, and the restriction of $\E$ to $\bar{\D}_{EXT}$ 
yields a singularity free SAV metric that satisfies all of the requisite 
matching conditions.  Again, we obtain a global solution that some 
relativists would regard as asymptotically flat.  However, in this case, 
in a neighborhood of spatial infinity, though 
\begin{equation}
\text{the limit $\nu_{NUT}$ of } \sqrt{z^{2}+\rho^{2}} \chi(z,\rho)
\text{ as } \sqrt{z^{2}+\rho^{2}} \rightarrow \infty
\end{equation}
exists, it is not zero.  For this reason, the spacetime is not asymptotically
flat in an orthodox sense.  No example of this case is known, and it would
be a shock if an example were found.

\item
Suppose that $\bar{\Sigma}_{ex}$ is a proper subspace of $\bar{\Sigma}_{EXT}$,
and $\Pi(\tau)$ has no holomorphic extension to $\hat{\Sigma} \cup 
\Sigma_{EXT}$ (i.e., to a domain that covers $\bar{\Sigma}_{EXT}$),  
regardless of the choice of $\bK_{4}$.  Then 
$\E$ has an analytic extension to a domain which covers $\bar{\Sigma}_{ex}$ 
and satisfies $\E(z,-\rho) = \E(z,\rho)$.  The restriction of $\E$ to 
$\bar{\D}_{ex}$ yields a singularity-free SAV region that envelopes the
fluid body and satisfies all requisite matching conditions.  However, the 
full analytic continuation of $\E(z,\rho)$ will have at least one singularity 
on $\bar{\D}_{EXT}$ or at spatial infinity and will, therefore, not furnish 
a global solution without singularities.  The Wahlquist solution may be in 
this case.

\item
Suppose that the holomorphic monodromy matrix $\Pi(\tau)$ does not exist
[i.e., the conditions (ii) in Sec.\ \ref{mat}C do not hold].  Then,
either our matching formalism is not applicable to the given SAPF or there
exists no SAV envelope that is free of singularities (including cusps) and
that matches the given SAPF.  When our matching formalism is applicable,
the matching SAV envelope of the given SAPF is unique (in the sense that
any two matching SAPF envelopes will have the same full analytic continuation).
If our matching formalism is not applicable, then there may exist two or more
matching SAPF envelopes with different full analytic continuations.
Criteria that would tell us [before computing $\Q(\lambda,\tau)$] when our
formalism is applicable remain to be discovered.
\end{romanlist}
All of the conclusions that have been given above in (i) through (iv) follow 
from previous work by the authors.\cite{he1,he2,he3}  

\subsection{Constructing Exterior Solution from $\Pi(\tau)$} 

If it is desired actually to construct the SAV solution that matches 
the given SAPF solution, one proceeds to identify a Kinnersley-Chitre 
transformation matrix $v(\tau)$ such that\footnote{Some choices of the
factorization $\Pi(\tau) = v(\tau) v(\tau)^{T}$ may only be applicable
to a domain $\hat{\Sigma} - \Sigma_{cut}$, where $\Sigma_{cut}^{*} = 
\Sigma_{cut}$ and is a union of cuts in $\hat{\Sigma}$.  Each of these
cuts crosses the real axis and its endpoints are ($\tau$-independent)
isolated branch points of $v(\tau)$.  However, these branch points lead
to no spacetime singularities, since the Ernst potential $\E(z,\rho)$
that is determined from $v(\tau)$ will have an analytic continuation
that covers $\bar{\D}_{ex}$.  In summary, there may be singularities
of $v(\tau)$ that are not singularities of $\Pi(\tau)$; and it is only
the singularities of $\Pi(\tau)$ that count.}
\begin{equation}
\Pi(\tau) = v(\tau)v(\tau)^{T}, 
\quad [v(\tau)]^{*} = v(\tau^{*}),
\text{ and } \det{v(\tau)} = 1.
\label{53}
\end{equation}
Clearly, the matrix $v(\tau)$ is defined by Eq.\ (\ref{53}) only up to 
a transformation
\begin{equation}
v(\tau) \rightarrow v(\tau)B(\tau),
\end{equation}
where $B(\tau)$ is an orthogonal $2 \times 2$ matrix that satisfies
$\det{B(\tau)}=1$ and $B^{*}(\tau)=B(\tau)$.  The choice of $B(\tau)$
has no effect upon the SAV solution.  The axis values of the $\E$
potential on the right side of the spinning spheroid are given by 
\begin{eqnarray}
\E(z,0) & = & \frac{1 + i \Pi_{34}(z)}{\Pi_{44}(z)} \nonumber \\
& = & \frac{v_{33}(z)+iv_{34}(z)}{-iv_{43}(z)+v_{44}(z)}
\text{ for all } z \ge z(0),
\label{axis}
\end{eqnarray}
from which, as is well known, the SAV spacetime can be constructed
by many modern methods that are based upon Riemann-Hilbert problems.

For example, our HHP corresponding to $v(\tau)$ can be solved in two 
successive steps\cite{he3}, each of which involves well known mathematics.  
In the first step, which requires only that we compute a definite integral 
with a given integrand, Minkowski space is transformed into a Weyl static 
spacetime.  In the second step, which requires that we solve an ordinary 
Fredholm equation of the second kind with a given kernel and a given 
inhomogeneous term, the Weyl static spacetime is transformed into the 
final SAV spacetime. 

\setcounter{equation}{0}
\section{Simple Illustrative Examples\label{examples}} 

We shall illustrate all stages of the $\Pi$-matrix method with 
examples.  In this discussion we shall employ geometrical units;
i.e., $G=1$ and $c=1$.  Moreover, the time coordinate will be scale 
and the arbitrary constant in $\chi$ selected so that 
$\Q_{0} := \Q(0,\tau) = I$ in Eqs.\ (\ref{3.qIC}) and (\ref{3.PI.2}).
In both cases the matching data $\rho(\lambda)$, $f(\lambda)$ and
$\omega(\lambda)$ are obtained directly from the metric, while simple
calculations yield the matching data $z(\lambda)$, $\chi(\lambda)$
and $\phi(\lambda)$.  

\subsection{The Schwarzschild Interior Solution}

The first example is the Schwarzschild interior 
solution\cite{tolman},
\begin{equation}
ds^{2} = \frac{dr^{2}}{1-2Mr^{2}/R^{3}} + r^{2}\left(
d\theta^{2} + \sin^{2}\theta d\varphi^{2}\right) - e^{2\psi} dt^{2},
\end{equation}
where $0 \le r \le R$ and\footnote{Note that we have scale the time
coordinate $t$ so that $\psi(\x)=0$ at $r=R$.} 
\begin{equation}
\psi(\x) = \ln \left[ \frac{3}{2} - \frac{1}{2} k^{-1}
(1-2Mr^{2}/R^{3})^{1/2}\right], \quad k := (1-2M/R)^{1/2}.
\end{equation}
In this chart, the pressure is 
given by
\begin{equation}
p(\x) = \frac{M}{(4\pi/3)R^{3}} \left(
\frac{\sqrt{1-2Mr^{2}/R^{3}} - \sqrt{1-2M/R}}
{3\sqrt{1-2M/R} - \sqrt{1-2Mr^{2}/R^{3}}} \right),
\end{equation}
which is independent of $\theta$ and which vanishes when $r=R$.  

Thus, for example, in the case of the Schwarzschild interior solution, 
the fields $P(\x)$, $f(\x)$, $\omega(\x)$ and $\chi(\x)$ are defined 
everywhere, and we find
\begin{equation}
P(\x) = r e^{\psi(\x)} \sin\theta, \quad
f(\x) = e^{2\psi(\x)}, \quad
\omega(\x) = 0, \quad \chi(\x) = 0,
\end{equation}
while
\begin{eqnarray}
\star dP(\x) & = & 
- \left[ e^{\psi} (1-2Mr^{2}/R^{3})^{1/2} + k^{-1} Mr^{2}/R^{3}\right]
r \sin\theta d\theta \nonumber \\
& & \mbox{} + e^{\psi} (1-2Mr^{2}/R^{3})^{-1/2} \cos{\theta} dr, \\
\eta(\x) & = & - k^{-1} M (r/R)^{3} \sin\theta d\theta,
\end{eqnarray}
where we have used the relations
\begin{equation}
\star dr = -\sqrt{r^{2}-2Mr^{4}/R^{3}} d\theta, \quad
\star d\theta = dr/\sqrt{r^{2}-2Mr^{4}/R^{3}}.
\end{equation}
On the zero pressure surface $r=R$, we obtain the matching data
\begin{equation}
\framebox{$
\begin{array}{l}
\text{Schwarzschild Interior Solution:} \\
z(\lambda) = k^{-1} (R-M) \cos\lambda, \quad
\rho(\lambda) = R \sin\lambda, \\
f(\lambda) = 1, \quad 
\omega(\lambda) = 0, \quad
\chi(\lambda) = 0, \quad
\phi(\lambda) = k^{-1} M \cos\lambda + \mathrm{const}.
\end{array}
$}
\end{equation}
The value of the constant may be chosen at will.
Since only $\dot{\phi}(\lambda)$ will be used, there is no need
to be more specific.

Incidentally, in the case of the Schwarzschild interior solution,
one can combine our expressions for $z(\lambda)$ and $\rho(\lambda)$
in the neat formula
\begin{equation}
z(\lambda) + i \rho(\lambda) = k^{-1} M \cosh(\beta + i\lambda), \quad
\cosh\beta := (R-M)/M,
\label{zirho}
\end{equation}
which has a holomorphic extension
\begin{equation}
z(\x) + i \rho(\x) = k^{-1} M \cosh(\beta + i\theta), \quad
\cosh\beta := (r-M)/M,
\end{equation}
where the fields $z(\x)$ and $\rho(\x)$ satisfy 
\begin{equation}
dz(\x) = \star d\rho(\x) \text{ and } d\rho(\x) = - \star dz(\x).
\end{equation}
While it is possible to employ a harmonic chart $(z,\rho)$ to 
describe the solution, within the fluid the field $\rho(\x)$ is 
{\em not\/} the same thing as the field $P(\x)$, which is quite 
complicated when expressed in terms of $z$ and $\rho$.

In the Weyl case, where $\omega(\x)=\chi(\x)=0$, Eq.\ (\ref{diffeq}) 
reduces to
\begin{equation}
\frac{\partial \Q(\lambda,\tau)}{\partial \lambda}
= - \frac{(\tau-z(\lambda))\dot{\psi}(\lambda) 
+ \dot{\phi}(\lambda)}{\mu(\x(\lambda),\tau)}
\sigma_{3} \Q(\lambda,\tau).
\end{equation}
Introducing $\Psi(\lambda,\tau)$ such that 
\begin{equation}
\Q(\lambda,\tau) = e^{-\Psi(\lambda,\tau)\sigma_{3}}, \quad \Psi(0,\tau) = 0,
\label{Psi2Q}
\end{equation}
we may write
\begin{equation}
\frac{\partial \Psi(\lambda,\tau)}{\partial \lambda}
= \frac{(\tau-z(\lambda))\dot{\psi}(\lambda) 
+ \dot{\phi}(\lambda)}{\mu(\x(\lambda),\tau)}.
\end{equation}
Thus, in the case of the Schwarzschild interior solution, we have
\begin{equation}
\frac{\partial \Psi(\lambda,\tau)}{\partial \lambda} = 
\frac{- k^{-1} M \sin\lambda}{[(\tau-z(\lambda))^{2} 
+ \rho(\lambda)^{2}]^{1/2}},
\end{equation}
and hence, 
\begin{equation}
\Psi(\lambda,\tau) = \ln \left[
\frac{\left(\frac{R-M}{M}\right)\tau - \left(\frac{M}{R-M}\right) z(\lambda)
- \mu(\x(\lambda),\tau)}
{\left(\frac{R-2M}{M}\right)\left(\tau+\left(\frac{M}{R-M}\right)z(0)\right)} 
\right].
\end{equation}
The function $\Q(\lambda,\tau)$ is given by Eq.\ (\ref{Psi2Q}), and
\begin{equation}
\Q'(\lambda,\tau) = e^{-\Psi'(\lambda,\tau)\sigma_{3}},
\end{equation}
where
\begin{equation}
\Psi'(\lambda,\tau) = \ln \left[
\frac{\left(\frac{R-M}{M}\right)\tau - \left(\frac{M}{R-M}\right) z(\lambda)
+ \mu(\x(\lambda),\tau)}
{\left(\frac{R-2M}{M}\right)\left(\tau+\left(\frac{M}{R-M}\right)z(0)\right)} 
\right].
\end{equation}
Hence
\begin{equation}
\Pi(\tau) := - J \Q'(0,\tau) J = e^{2\xi(\tau)\sigma_{3}},
\end{equation}
where
\begin{equation}
\xi(\tau) := -\ln{k} + \frac{1}{2} \ln \left( \frac{k\tau-M}{k\tau+M} \right).
\end{equation}
Therefore, we can select the K-C transformation matrix
\begin{equation}
v(\tau) = e^{\xi(\tau)\sigma_{3}}
= \left( \begin{array}{cc}
k^{-1} & 0 \\ 0 & k
\end{array} \right) \left( \begin{array}{cc}
\left(\frac{k\tau-M}{k\tau+M}\right)^{\frac{1}{2}} & 0 \\
0 & \left(\frac{k\tau+M}{k\tau-M}\right)^{\frac{1}{2}}
\end{array} \right).
\end{equation}
By Eq.\ (\ref{axis}) the axial values of the $\E$-potential are given by
\begin{equation}
k^{2} \E(z,0) = \frac{kz-M}{kz+M}, \quad kz \ge R - M. 
\end{equation}
The factors $k^{2}$ and $k$ can be suppressed by rescaling the time
coordinate $t$ and the spectral parameter $\tau$.
We then obtain the well known axis values of the
exterior Schwarzschild solution.  The values of $\E(z,\rho)$ for
$\rho > 0$ are obtained by solving the authors' HHP or one of the 
other well known methods. 

\subsection{A Winicour Dust Solution}

A somewhat more complicated example is provided by a Winicour dust 
metric\cite{winicour}
\begin{equation}
ds^{2} = e^{-b^{2}\rho^{2}/4} (dz^{2} + d\rho^{2}) 
+ \rho^{2} d\varphi^{2} 
- [dt + (b\rho^{2}/2) d\varphi]^{2},
\end{equation}
for which the pressure vanishes everywhere, while the energy density
is given by\footnote{We employ units such that $c=G=1$.}
\begin{equation}
\epsilon(z,\rho) = \frac{1}{8\pi}b^{2} e^{b^{2}\rho^{2}/4}.
\end{equation}
Because the pressure vanishes everywhere, one can select any convenient 
matching surface; for example,
\begin{equation}
z = R \cos\lambda, \quad \rho = R \sin\lambda,
\label{zps}
\end{equation}
where $0 \le \lambda \le \pi$.  Note that the requirement $g_{33}>0$ 
yields $(bR/2)^{2} < 1$.

In the case of the dust metric, $Z(\x)$, $P(\x)$, $f(\x)$, $\omega(\x)$ 
and $\chi(\x)$ are defined throughout the chart, and we find
\begin{equation}
Z(\x) = z, \quad P(\x) = \rho, \quad f(\x) = 1, \quad 
\omega(\x) = -\frac{1}{2}b\rho^{2}, \quad \chi(\x) = b(z-R),
\end{equation}
while
\begin{equation}
\eta = \frac{1}{4} b^{2} \rho^{2} dz.
\end{equation}
On the selected zero pressure surface (\ref{zps}), we obtain the matching
data
\begin{equation}
\framebox{$
\begin{array}{l}
\text{Winicour Dust Metric:} \\
z(\lambda) = R \cos\lambda, \quad \rho(\lambda) = R \sin\lambda, \\
f(\lambda) = 1, \quad 
\omega(\lambda) = - \frac{1}{2} bR^{2} \sin^{2}\lambda, \\
\chi(\lambda) = b R (\cos\lambda-1), \quad 
\phi(\lambda) = \frac{1}{4} b^{2} R^{2} \left(
\cos\lambda - \frac{1}{3} \cos^{3}\lambda - \frac{2}{3}\right).
\end{array}
$}
\end{equation}
When $f(\lambda)$ is independent of $\lambda$, it is obvious from Eqs.\ 
(\ref{2.26}) and (\ref{Delta}) that those terms in $\Delta(\x,\tau)$
that are proportional to the matrix $\sigma_{3}$ all vanish.\footnote{For 
this reason it is not really necessary to evaluate $\phi(\lambda)$.}
We find that Eq.\ (\ref{diffeq}) assumes the form 
\begin{eqnarray}
\frac{\partial \Q(\lambda,\tau)}{\partial\lambda} & = &
\frac{b}{2} R \sin\lambda \left[ \frac{\tau}{m(\lambda,\tau)}
\sigma_{1} + J \right] \Q(\lambda,\tau), 
\label{2.2.14} \\ \Q(0,\tau) & = & I, 
\nonumber
\end{eqnarray}
where 
\begin{equation}
m(\lambda,\tau) := [\tau^{2}-2\tau R\cos\lambda + R^{2}]^{1/2}.
\label{2.2.21}
\end{equation}

Consider the following differential equation in the complex plane:\footnote{
For fixed $\tau$, each matrix element of $\F$ satisfies the fourth order 
equation $\Box^{2}\F + \beta^{2}\F = 0$ with $\Box := \partial^{2}/\partial
\zeta^{2} - \beta^{2} (\tau^{2} - \zeta^{2})$ and $\beta := b/(2\tau)$.  
We do not know if this differential equation or the second order equations 
$\Box\F \pm i \beta \F = 0$ have been the subject of study elsewhere.}
\begin{equation}
\frac{\partial \F(\zeta,\tau)}{\partial\zeta} =  \frac{b}{2}
\left( \sigma_{1} + \frac{\zeta}{\tau} J \right) \F(\zeta,\tau)
\label{2.2.28}
\end{equation}
such that
\begin{equation}
\dom{\F} = \{ (\zeta,\tau) \in C^{2}: \zeta \ne \infty \text{ and }
\tau \notin \{0,\infty\} \}
\label{2.2.29}
\end{equation}
and 
\begin{equation}
\F(\tau-R,\tau) = I.
\label{2.2.30}
\end{equation}
If $\F$ is the function defined by the above Eqs.\ (\ref{2.2.28}) to 
(\ref{2.2.30}), then it is easy to prove with the aid of 
\begin{equation}
\frac{\partial m(\lambda,\tau)}{\partial \lambda} = R\sin\lambda 
\frac{\tau}{m(\lambda,\tau)}
\end{equation}
and
\begin{equation}
m(0,\tau) = \tau - R
\end{equation}
that
\begin{equation}
\Q(\lambda,\tau) := \F(m(\lambda,\tau),\tau)
\label{2.2.31}
\end{equation}
is the solution of Eqs.\ (\ref{2.2.14}) over the domain
\begin{equation}
\{(\lambda,\tau):0 \le \lambda \le \pi \text{ and } \tau \in C - 
\bar{K}(\x(\lambda)) - \{0,\infty\} \}.
\label{2.2.32}
\end{equation}

The triad of equations (\ref{2.2.28}), (\ref{2.2.29}) and 
(\ref{2.2.30}) is collectively equivalent to Eq.\ (\ref{2.2.29}) 
taken together with the following integral equation:
\begin{equation}
\F(\zeta,\tau) = I + \frac{b}{2} \int_{\tau-R}^{\zeta} d\zeta'
\left( \sigma_{1} + \frac{\zeta'}{\tau} J \right) \F(\zeta',\tau)
\end{equation}
or, equivalently,
\begin{equation}
\G(\eta,\tau) = I + \frac{b}{2} \int_{0}^{\eta} d\eta'
\left( \gamma(\tau) + J \frac{\eta'}{\tau} \right) \G(\eta',\tau),
\label{2.2.34}
\end{equation}
where
\begin{equation}
\gamma(\tau) := \sigma_{1} + J \left( \frac{\tau-R}{\tau} \right)
\end{equation}
and
\begin{equation}
\eta := \zeta - (\tau-R) \text{ and } \G(\eta,\tau) := \F(\eta+\tau-R,\tau).
\label{2.2.35}
\end{equation}
The solution of Eq.\ (\ref{2.2.34}) is given by the infinite series
\begin{equation}
\G(\eta,\tau) = \sum_{n=0}^{\infty} \left(\frac{b}{2}\right)^{n}
\G_{n}(\eta,\tau),
\label{2.2.36}
\end{equation}
where
\begin{equation}
\G_{0}(\eta,\tau) = I
\end{equation}
and, for all $n \ge 0$,
\begin{equation}
\G_{n+1}(\eta,\tau) = \int_{0}^{\eta} d\eta'
\left[ \gamma(\tau) + J \frac{\eta'}{\tau} \right] \G_{n}(\eta',\tau).
\end{equation}
By mathematical induction one proves that
\begin{equation}
\G_{n}(\eta,\tau) = \frac{\eta^{n}}{n!} g_{n}(\eta,\tau),
\label{4.SOL.1}
\end{equation}
where
\begin{equation}
g_{n}(\eta,\tau) = \sum_{k=0}^{n} g_{nk}(\tau) (\eta/\tau)^{k}
\label{4.SOL.2}
\end{equation}
and the coefficients $g_{nk}(\tau)$ are to be computed from the
recursion relation
\begin{equation}
g_{n+1,k}(\tau) = \left(\frac{n+1}{n+1+k}\right)
\left[\gamma(\tau)g_{nk}(\tau)+J g_{n,k-1}(\tau)\right]
\label{4.SOL.3}
\end{equation}
and the conditions
\begin{equation}
g_{00}(\tau) = I, \; g_{nk}(\tau) = 0 \text{ if $k<0$ and if $k>n$.}
\label{4.SOL.4}
\end{equation}
Thus, $g_{0}(\eta,\tau)=I$,
\begin{equation}
g_{1}(\eta,\tau) = \gamma(\tau) + J \frac{\eta}{2\tau}
\label{g1}
\end{equation}
and
\begin{equation}
g_{2}(\eta,\tau) = \gamma(\tau)^{2} + 
\left( J \gamma(\tau) + \frac{1}{2} \gamma(\tau) J \right) 
\frac{2\eta}{3\tau} - I \frac{\eta^{2}}{4\tau^{2}},
\label{g2}
\end{equation}
where
\begin{equation}
\gamma(\tau)^{2} = I \left[ 1 - \left(\frac{\tau-R}{\tau}\right)^{2} \right]
\end{equation}
and, since $J \sigma_{1} = \sigma_{3}$,
\begin{equation}
J \gamma(\tau) + \frac{1}{2} \gamma(\tau) J = \frac{1}{2} \sigma_{3}
- \frac{3}{2} I \left( \frac{\tau-R}{\tau} \right).
\end{equation}
One further proves by mathematical induction that
\begin{equation}
g_{nk}(\tau) \text{ is a holomorphic function of $\tau$ throughout }
C - \{0\}
\label{4.Gnk.hol1}
\end{equation}
and
\begin{equation}
\begin{array}{l}
\tau^{n} g_{nk}(\tau) \text{ is a holomorphic function of } \tau \\
\text{throughout } C - \{\infty\} \text{ and has the value } \\
(-JR)^{n} \delta_{k0} \text{ at } \tau=0.
\end{array}
\label{4.Gnk.hol2}
\end{equation}

From Eqs.\ (\ref{2.2.31}) and (\ref{2.2.35}),
\begin{equation}
\Q(\lambda,\tau) = \F(m(\lambda,\tau),\tau) = \G(m(\lambda,\tau)-(\tau-R),\tau)
\label{2.2.48}
\end{equation}
for all $(\lambda,\tau)$ in the domain (\ref{2.2.32}).  From Eqs.\ 
(\ref{2.2.36}), (\ref{4.SOL.1}), (\ref{4.SOL.2}), (\ref{4.Gnk.hol1}),
(\ref{4.Gnk.hol2}) and the existence of the limits
\begin{equation}
\lim_{\tau\rightarrow\infty} \left[ m(\lambda,\tau) - (\tau-R) \right] =  
R(1-\cos\lambda)
\label{lim1}
\end{equation}
and 
\begin{equation}
\lim_{\tau \rightarrow 0} 
\left[ \frac{m(\lambda,\tau)-(\tau-R)}{\tau} \right] = 
- \left( 1 - \cos\lambda \right),
\label{lim2}
\end{equation}
it follows as expected that
\begin{equation}
\begin{array}{l}
\text{for each } \lambda \in [0,\pi], \F(m(\lambda,\tau),\tau) 
\text{ has a } \\
\text{holomorphic extension to all of } \\
C - \bar{K}(\x(\lambda)),
\end{array}
\label{2.2.49}
\end{equation}
and the equality (\ref{2.2.48}) holds for all $(\lambda,\tau)$
in the domain
\begin{equation}
\dom{\Q(\lambda,\tau)} := 
\{(\lambda,\tau):0 \le \lambda \le \pi \text{ and } 
\tau \in C - \bar{K}(\x(\lambda))\}.
\end{equation}
Incidentally, the reader can use Eqs.\ (\ref{lim1}) and (\ref{lim2})
to compute
\begin{eqnarray}
\Q(\lambda,\infty) & = & \exp{\left[\frac{b}{2}R(1-\cos\lambda)
(\sigma_{1}+J)\right]}
\nonumber \\
& = & I + bR (1-\cos\lambda) \left( \begin{array}{cc}
0 & 1 \\ 0 & 0
\end{array} \right), \\
\Q(\lambda,0) & = & \exp{\left[\frac{b}{2} R (1-\cos\lambda) J\right]}.
\end{eqnarray}

Since $\Q_{0} := \Q(0,\tau) = I$ in Eq.\ (\ref{3.PI.2}), the $Pi$-matrix
is given by
\begin{equation}
\Pi(\tau) = - J \G(-2(\tau-R),\tau) J.
\end{equation}
So, from Eqs.\ (\ref{2.2.36}), (\ref{4.SOL.1}) and (\ref{4.SOL.2}),
\begin{equation}
\Pi(\tau) = -J \sum_{n=0}^{\infty} \frac{[-b\tau(1-R/\tau)]^{n}}{n!}
g_{n}(-2(\tau-R),\tau) J
\end{equation}
and
\begin{equation}
g_{n}(-2(\tau-R),\tau) = \sum_{k=0}^{n} g_{nk}(\tau)
\left[-2(1-R/\tau)\right]^{k}.
\end{equation}
Equations (\ref{g1}) and (\ref{g2}) imply that
$g_{1}(-2(\tau-R),\tau)$ and $g_{2}(-2(\tau-R),\tau)$ are not zero
at $\tau=\infty$.  Further calculations reveal that $g_{n}(-2(\tau-R),\tau)$
is not zero at $\tau=\infty$ for all $n \le 10$.  Granting that this
continues to be true for all values of $n$, we may conclude that $\Pi(\tau)$
has an isolated essential singularity at $\tau=\infty$, and, therefore,
the SAV that matches the dust metric at the boundary we have been
considering is not asymptotically flat.  A proof of this conclusion,
however likely it seems, remains to be found.

\setcounter{equation}{0}
\section{Generalizations\label{con}} 

The $\Pi$-matrix method can be generalized in a number of respects.  The 
stationary axisymmetric source need not be of spheroidal shape.  Toroidal 
sources, or multiple spheroidal sources spinning on a common axis, would be 
interesting.  The source need not even be a perfect fluid.  Stationary 
axisymmetric charge and current density may be involved, with resulting 
stationary axisymmetric electromagnetic fields.  In principle, all such 
``joining problems'' can be handled by an extended $\Pi$-matrix approach.  
Once the axis values of the the complex potentials $\E$ and $\Phi$ have 
been deduced from $\Pi(\tau)$, the exterior electrovac fields can be 
constructed by solving the electrovac version of the authors' HHP.  In 
this case the calculational methods devised by Alekseev\cite{alekseev}
and Sibgatullin\cite{sib} are germaine.

We have in this first exposition of the $\Pi$-matrix method avoided the 
mathematical complications that such generalizations would entail, hoping 
that this would make it easier for the reader to appreciate the general 
idea behind this approach.

\section*{Acknowledgement}
Research supported in part by grant PHY-98-00091 from
the National Science Foundation to FJE Enterprises.

\newpage
\begin{figure} 
\font\thinlinefont=cmr5
\begingroup\makeatletter\ifx\SetFigFont\undefined%
\gdef\SetFigFont#1#2#3#4#5{%
  \reset@font\fontsize{#1}{#2pt}%
  \fontfamily{#3}\fontseries{#4}\fontshape{#5}%
  \selectfont}%
\fi\endgroup%
\mbox{\beginpicture
\setcoordinatesystem units <1.00000cm,1.00000cm>
\unitlength=1.00000cm
\linethickness=1pt
\setplotsymbol ({\makebox(0,0)[l]{\tencirc\symbol{'160}}})
\setshadesymbol ({\thinlinefont .})
\setlinear
%
%
\linethickness= 0.500pt
\setplotsymbol ({\thinlinefont .})
\ellipticalarc axes ratio  5.080:5.080  180 degrees 
	from 12.700 19.050 center at  7.620 19.050
%
%
\linethickness= 0.500pt
\setplotsymbol ({\thinlinefont .})
\ellipticalarc axes ratio  2.540:2.540  180 degrees 
	from 10.160 19.050 center at  7.620 19.050
%
%
\linethickness= 0.500pt
\setplotsymbol ({\thinlinefont .})
\putrule from  2.540 19.050 to 12.700 19.050
%
%
\put{\SetFigFont{12}{14.4}{\rmdefault}{\mddefault}{\updefault}$\M^{2}_{ex}$} [lB] at  7.144 22.543
%
%
\put{\SetFigFont{12}{14.4}{\rmdefault}{\mddefault}{\updefault}$\M^{2}_{in}$} [lB] at  7.144 19.844
%
%
\put{\SetFigFont{12}{14.4}{\rmdefault}{\mddefault}{\updefault}$\M^{1}_{0}$} [lB] at  9.1 21.0 
%
%
\put{\SetFigFont{12}{14.4}{\rmdefault}{\mddefault}{\updefault}$\M^{1}_{ax}$} [lB] at  3.493 19.050
%
%
\put{\SetFigFont{12}{14.4}{\rmdefault}{\mddefault}{\updefault}$\M^{1}_{ax}$} [lB] at 11.113 19.050
%
%
\put{\SetFigFont{12}{14.4}{\rmdefault}{\mddefault}{\updefault}$\M^{1}_{ax}$} [lB] at  7.144 19.050
\linethickness=0pt
\putrectangle corners at  2.515 24.145 and 12.725 13.955
\endpicture}
\caption{Subspaces of the topological space $\M^{2}$}
\end{figure}

\end{document}